\documentclass[10pt,twocolumn]{article}

\usepackage[T1]{fontenc}
\usepackage{lmodern}
\usepackage{microtype}
\usepackage{hyperref}
\usepackage{graphicx}
\usepackage[section]{placeins}
\usepackage{booktabs}
\usepackage{multirow}
\usepackage{amsmath}
\title{BEDCrypt: Privacy-preserving interval analytics with homomorphic encryption}
\author{Kimon Antonios Provatas \and Ilias Georgakopoulos-Soares}
\date{}

\begin{document}
\maketitle

\footnotetext[1]{* Corresponding author: \texttt{ilias@austin.utexas.edu}}

\begin{abstract}
\textbf{Motivation.} Genomic data and derived interval datasets can carry sensitive information, and the analysis itself can reveal an analyst's intent. As genomic workloads are increasingly outsourced to third-party infrastructure, there is a need for privacy-preserving technologies that protect both the data and the queried loci.

\textbf{Results.} We present \textsc{BEDCrypt}, a privacy-preserving system for genomic interval analytics based on homomorphic encryption in an honest-but-curious server setting. The server operates only on encrypted data and returns encrypted answers that the client decrypts locally, enabling core functionalities such as coverage summaries, interval intersections, proximity (window-style) queries, and set-similarity statistics, without revealing plaintext intervals or query genomic locations to the server.
\end{abstract}

\section{Introduction}
\label{sec:introduction}

\phantomsection\label{sec:implementation}
\phantomsection\label{sec:future-work}

Interval-based primitives such as coverage summaries, overlap and proximity queries, and set-similarity statistics are foundational to modern genomics workflows. They appear throughout quality control, annotation, and comparative analyses, and are widely accessed through tools such as \texttt{bedtools}~\cite{quinlan2010bedtools}.

However, genomic coordinates and derived interval tracks can reveal sensitive information, and the analyst's queries can themselves expose research intent. As practitioners increasingly rely on shared clusters and cloud platforms, there is a growing need for systems that enable genomic interval analytics while keeping both the underlying data and the queried loci private.

Homomorphic encryption (HE) provides one principled path to this goal by enabling computation over ciphertexts, so that an untrusted server can process encrypted data and return encrypted results for client-side decryption~\cite{gentry2009fhe,brakerski2012fv,fan2012some,seallib}. Recent surveys position HE as a core privacy-enhancing technology for biomedical data science~\cite{cho2024pets}. Prior work has successfully applied HE to several genomics workloads, including private variant presence queries, encrypted sequence search (e.g., we previously introduced private k-mer search in \textsc{KmerCrypt})~\cite{kmercrypt2025}, and large-scale statistical genetics pipelines such as GWAS~\cite{sim2020gwashe,blatt2020securegwas}, polygenic risk scores~\cite{knight2026heprs}, and privacy-preserving genotype imputation~\cite{gursoy2022pimpute}. However, these methods largely focus on fixed-position variant vectors or sequence-comparison tasks rather than the interval-algebra primitives used in BED-style workflows, motivating HE-friendly constructions specialized for coverage, intersection, and proximity queries. Yet applying HE to interval algebra is challenging because interval operations are naturally data-dependent and are often implemented via index traversals and branching.

Related lines of work in encrypted query processing illustrate alternative design points. CryptDB~\cite{popa2011cryptdb} targets SQL analytics over relational databases via a proxy that rewrites queries over adjustable encryption layers (including property-preserving encryption), rather than performing homomorphic computation on an HE-encrypted dataset. Subsequent work has shown that property-preserving techniques can admit powerful inference attacks~\cite{naveed2015inference}, and the underlying primitives (e.g., order-preserving encryption (OPE)~\cite{boldyreva2009ope} and order-revealing encryption (ORE)~\cite{boneh2015ore,lewi2016ore}) inherently expose order information---a particularly poor fit for genomic coordinates, which are ordered by construction. In contrast, \textsc{BEDCrypt} avoids server-side order comparisons and instead uses IND-CPA-secure HE to keep values confidential, while hiding the mapping from accesses to genomic loci via a client-secret permutation and relegating semantic reconstruction to the client.

At the level of interval primitives, W{\"u}ller et al.~\cite{wuller2017intervalops} study privacy-preserving interval operations using homomorphic encryption and secret sharing. BEDCrypt differs in its emphasis on a reusable, permuted SIMD-chunk database representation and an extraction-oriented protocol in which the server executes a fixed rotate-and-mask circuit to return only encrypted scalars needed for client-side reconstruction of \texttt{bedtools}-style outputs.

Secure set intersection has also been studied extensively, for example via OT-extension-based private set intersection (PSI)~\cite{pinkas2014psi} and high-throughput OPRF-based variants~\cite{kolesnikov2016oprf}. For genomic range-constrained intersection, specialized schemes such as PriRanGe have been proposed~\cite{prirange2023}. While PSI is a useful neighboring primitive for positioning ``intersection'' queries, BED-style interval analytics additionally requires range arithmetic (e.g., prefix-sum differences), overlap logic, and proximity constraints over an ordered coordinate line, which motivates different encodings and an extraction-oriented execution model.

Finally, prior HE-for-genomics systems have focused primarily on statistical genetics workloads such as encrypted GWAS pipelines~\cite{sim2020gwashe,blatt2020securegwas}, which involve substantial arithmetic over genotype/phenotype matrices. \textsc{BEDCrypt} targets a different core data type and access pattern---interval tracks and queried loci---and optimizes around selectively extracting encrypted boundary scalars needed for interval primitives (coverage/intersect/window/depth/Jaccard), rather than evaluating large linear-algebra models homomorphically.

We present \textsc{BEDCrypt}, a privacy-preserving system that supports common interval queries over encrypted genomic datasets in an honest-but-curious server setting. \textsc{BEDCrypt} adopts a \emph{permuted-chunk} architecture: the client first preprocesses the reference interval dataset into compact linear signals (e.g., prefix-sum style arrays for range queries and a depth array for point queries), partitions these signals into fixed-size SIMD batches, applies a client-secret random permutation over the batches, and encrypts each permuted batch under BFV~\cite{brakerski2012fv,fan2012some} (as implemented in Microsoft SEAL~\cite{seallib}).

Queries are answered via a \emph{blind extraction} protocol rather than genome-wide masked multiplication. For each scalar needed to reconstruct an output, the client sends only a permuted batch identifier and an intra-batch offset. The server performs a small, fixed sequence of homomorphic operations (rotation and masking) to extract that encrypted scalar and returns it. The client decrypts and combines extracted values locally to produce standard results for coverage summaries, interval intersections, proximity (window-style) queries, and set-similarity statistics, without revealing plaintext intervals or query locations to the server.

This design makes the server-side work proportional to the number of values requested, not the full genome representation, and it supports database reuse across multiple analyses once the encrypted permuted database is uploaded. The main residual leakage is through coarse metadata such as query volume and which permuted batches are accessed.

\section{Problem Definition and Threat Model}
\label{sec:threat-model}

\subsection{Interval analytics model}
\label{sec:interval-model}

We formalize the genomic coordinate space for a single chromosome as a discrete domain $x \in \{0,1,\dots,L-1\}$, and extend to whole genomes by treating chromosomes independently with their respective lengths.

\paragraph{Database intervals ($B$).}
The client designates one BED-like interval set $B = \{(s_j,e_j)\}_{j=1}^m$ as the reference database that is preprocessed once and hosted (in encrypted form) on the server.

\paragraph{Database preprocessing (logical, per-array).}
All preprocessing is performed in logical genomic order before any permutation.
From $B$ we build discrete arrays over the chromosome coordinate domain:
(i) \emph{Coverage prefix sum} $P_{\mathrm{cov}}(x)$, obtained by converting coverage to a binary array $I_B(x)$ and prefix-summing it, enabling covered-bases range queries via boundary differences;
(ii) \emph{Starts prefix sum} $P_{\mathrm{start}}(x)$, the prefix sum of an indicator that is 1 at each start coordinate of an interval in $B$;
(iii) \emph{Ends prefix sum} $P_{\mathrm{end}}(x)$, the (shifted) prefix sum of an indicator that is 1 at each end coordinate of an interval in $B$; and
(iv) \emph{Depth array} $D_B(x)$, where each entry equals the number of intervals in $B$ covering $x$.

\paragraph{Query intervals ($A$).}
For each analysis run, the client provides a query set $A = \{(s_i,e_i)\}_{i=1}^n$ whose loci must remain private from the server.

\paragraph{Query selectors and permutation (physical addressing).}
The logical database arrays are concatenated, partitioned into fixed-size chunks aligned with the HE batching capacity, and then permuted by a client-secret shuffle before encryption and upload.
Given a logical coordinate request (e.g., ``read $P_{\mathrm{cov}}(s)$''), the client computes the corresponding logical chunk index and intra-chunk offset, maps the chunk through the secret permutation to a physical chunk identifier, and emits a compact \emph{selector} describing (physical chunk id, slot offset).
The server uses these selectors only to perform blind extraction of encrypted scalars and cannot map them back to genome coordinates.

\paragraph{Supported query classes.}
Using the extracted boundary values, the client reconstructs:
(i) \emph{coverage summaries} (covered bases and overlap counts);
(ii) \emph{intersection existence} (any overlap vs. none);
(iii) \emph{proximity (window) queries} with margin $W$; and
(iv) \emph{set similarity (Jaccard)} aggregated over the query set.

\subsection{Threat model}
\label{sec:threat-model-semi-honest}

We assume an honest-but-curious (semi-honest) server that follows the prescribed protocol but attempts to learn information about the client's data and analytic intent from the transcript and its local state~\cite{goldreich2004foundations}.

\paragraph{Server observations.}
The server observes: (i) public cryptographic parameters and evaluation material (BFV parameters $N,t,q$ and the client's public and Galois keys); (ii) the encrypted, permuted database ciphertexts encoding $B$; and (iii) the \emph{permuted access pattern} induced by blind extraction, namely which physical chunk identifiers and intra-chunk slot offsets are requested.

\paragraph{Server non-observations.}
Under IND-CPA security of BFV, the server does not observe plaintext values (e.g., it cannot distinguish different coverage or depth values from ciphertexts)~\cite{brakerski2012fv,fan2012some}. Moreover, because chunks are permuted with a client-secret Fisher--Yates shuffle, the server cannot map a physical chunk identifier back to a logical genomic region (e.g., a specific chromosome coordinate range). Finally, because all supported features are decomposed into the same blind-extraction primitive, the server cannot directly infer whether a request corresponds to coverage, window/intersect, or Jaccard semantics.

\subsection{Leakage discussion}
\label{sec:leakage}

To avoid the overhead of full-fledged oblivious RAM (ORAM) while retaining practical performance, \textsc{BEDCrypt} accepts a coarse and explicitly modeled leakage profile~\cite{goldreich1996software,stefanov2013path}.

\paragraph{Leakage and protection.}
\textsc{bedCrypt} provides semantic security for the hosted database: under IND-CPA security of BFV, the server learns no plaintext information about the database intervals $B$ or the derived arrays it stores (coverage, starts, ends, depth)~\cite{brakerski2012fv,fan2012some}. However, \textsc{bedCrypt} does not implement a fully oblivious access mechanism (e.g., ORAM) for query processing~\cite{goldreich1996software,stefanov2013path}. The server learns coarse query-side metadata, including the total number of extracted scalars and which \emph{permuted} chunks are accessed (and how often) during a session. While the client-secret permutation prevents mapping these accesses to logical genomic loci, repeated or highly non-uniform query workloads may still yield statistical patterns about the query set $A$ through access frequencies, consistent with known leakage-abuse and inference attacks when access patterns or query repetitions are observable~\cite{cash2015leakage,islam2014range}. As a practical mitigation, the client may periodically refresh the hosted database by re-randomizing the chunk permutation and re-encrypting the permuted chunks, thereby rotating the physical layout visible to the server and reducing the long-term utility of frequency-based access-pattern observations across sessions.

\section{System Overview}
\label{sec:system-overview}

\begin{figure}[!htbp]
\centering
\includegraphics[width=\linewidth,keepaspectratio]{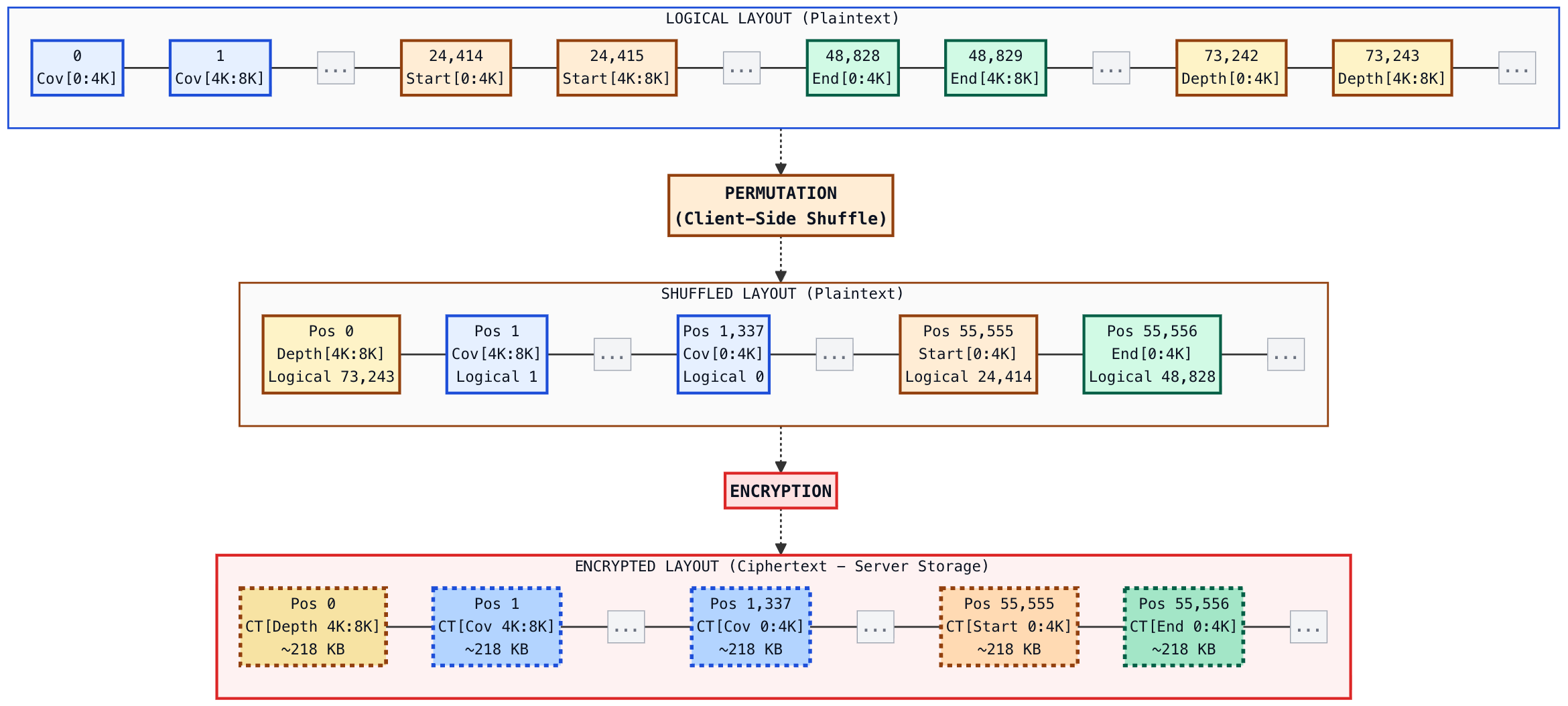}
\caption{\textbf{Encrypted database layout (high level).} The client converts the database intervals into linear plaintext arrays (coverage/start/end/depth), concatenates them into a logical vector, partitions the vector into fixed-size chunks (one BFV ciphertext per chunk), applies a client-secret permutation over chunk order, and encrypts each chunk for server-side storage. Query evaluation uses selectors that reference a permuted chunk identifier and an intra-chunk slot offset, enabling blind extraction without revealing logical genomic coordinates.}
\label{fig:db-layout}
\end{figure}

\begin{figure}[!htbp]
\centering
\includegraphics[width=\linewidth,height=0.68\textheight,keepaspectratio]{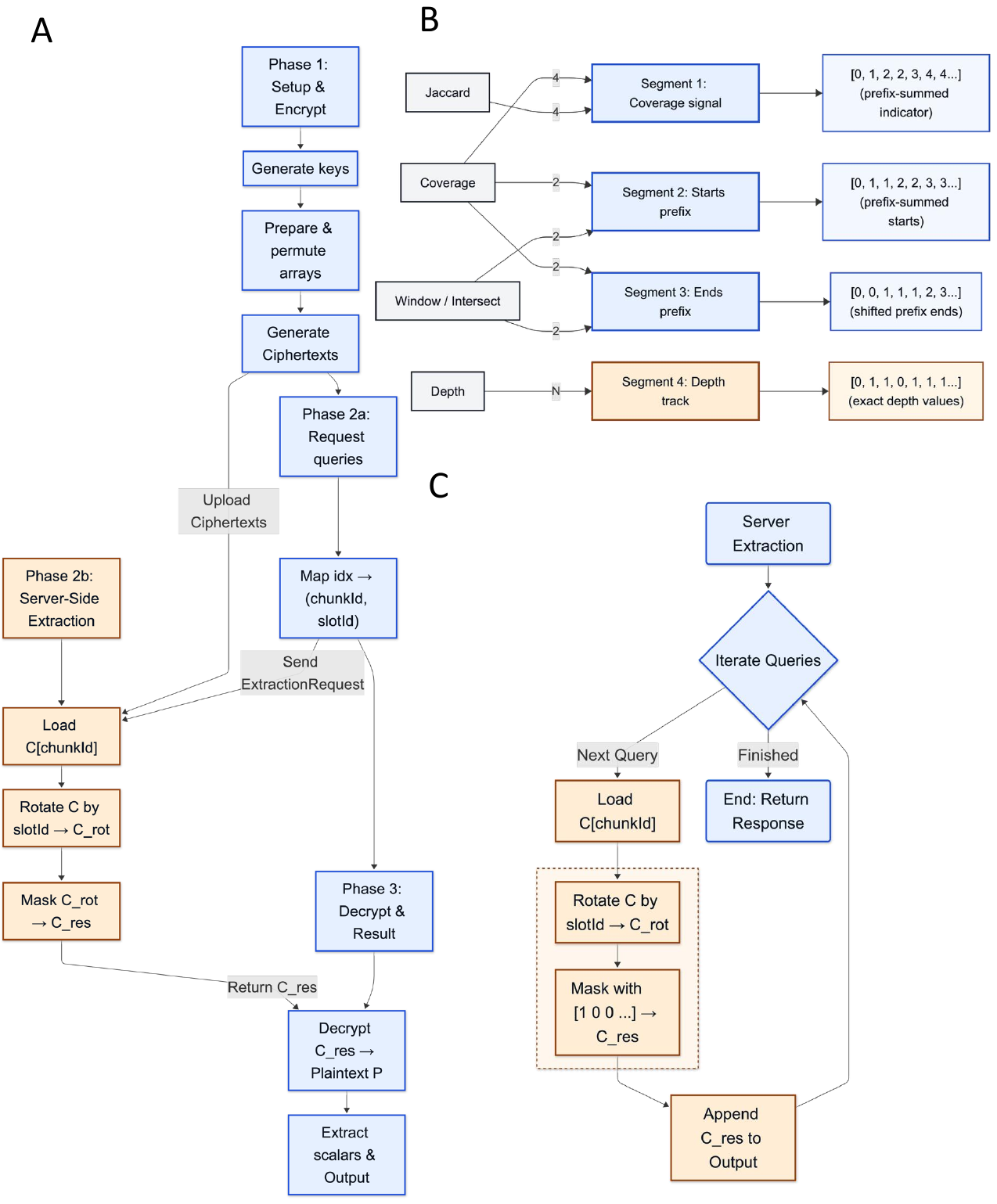}
\caption{\textbf{BEDCrypt System Design.} \textbf{(A) Architecture:} The end-to-end protocol separating client-side logic (setup, query compilation, and decryption) from server-side storage and processing. \textbf{(B) Functionality-to-Data Mapping:} The relationship between genomic operations (e.g., Coverage, Depth, Jaccard) and the underlying encrypted segments (S1--S4), defining which arrays are accessed and the specific scalar payloads (e.g., prefix sums vs. exact values) returned. \textbf{(C) Server-Side Computation:} The oblivious extraction kernel that processes requests by loading specific ciphertext chunks, rotating target values to the zero-th slot, and masking the remaining slots to isolate the result.}
\label{fig:system-design}
\end{figure}

\textsc{BEDCrypt} is implemented as a two-component system: a client executable that performs preprocessing, encryption, query compilation, and output reconstruction, and a server executable that hosts the encrypted database and executes query requests. Figure~\ref{fig:db-layout} summarizes the high-level encrypted database layout, and Fig.~\ref{fig:system-design} shows the end-to-end protocol and extraction kernel. The core operational idea is to (i) build and upload a reusable encrypted database once, then (ii) answer each subsequent analysis by extracting only the small set of encrypted scalars needed to reconstruct a \texttt{BEDTools}-style result, rather than scanning or reprocessing the full encrypted genome at query time (Fig.~\ref{fig:system-design}C). Figure~\ref{fig:system-design}B summarizes the mapping between supported interval operations and the underlying encrypted segments/arrays they access.

\FloatBarrier

\FloatBarrier

\subsection{Client-side responsibilities}
\label{sec:system-overview-client}

\paragraph{Database layout (high level).}
The client materializes the interval database $B$ into a small number of linear arrays over the chromosome coordinate space (coverage, start counts, end counts, and depth). These arrays are concatenated into a single logical vector, partitioned into fixed-size chunks aligned with the BFV batching capacity (one ciphertext per chunk), and then permuted by a client-secret shuffle before encryption. The server therefore stores a permuted sequence of ciphertext chunks that is reusable across queries, while the client maintains the permutation map and a lightweight manifest that maps each logical index (array type and coordinate) to a physical (chunk id, slot offset) selector.

\paragraph{Database build (one-time per $B$).}
The client parses the database BED file $B$ and constructs the interval-derived arrays used by the supported commands (prefix-sum arrays and a depth array). It concatenates these arrays into a single logical ``global'' vector per chromosome and records a layout manifest (segment offsets and lengths) so that later queries can refer to values by logical index.

\paragraph{SIMD packing and permuted chunks.}
The client partitions each logical global vector into fixed-size SIMD batches, where each batch fits into one BFV ciphertext (one chunk per ciphertext), enabling SIMD-style processing via batching in lattice-based HE~\cite{smart2014simd,fan2012some}. It then applies a client-side permutation to the chunk order and encrypts the permuted chunks (Fig.~\ref{fig:db-layout}). The resulting encrypted database is provisioned to the server once and can be reused across many query sessions.

\paragraph{Key material and protocol metadata.}
The client generates the public key and the rotation (Galois) keys needed for server-side slot rotations, and stores the secret key locally~\cite{halevi2014helib,seallib}. It also prepares a compact request description that specifies the chosen operation and references the encrypted database needed by the server.

\paragraph{Query compilation.}
For each query run, the client parses the query BED file $A$ and compiles the requested operation into a list of scalar reads from the logical global vectors. Each scalar read is translated into a physical address in the permuted representation (permuted chunk id and intra-chunk slot offset) and annotated with ordering information so decrypted scalars can be reassembled into the final output.

\paragraph{Result reconstruction.}
After receiving the server response, the client decrypts the returned ciphertext scalars, reorders them according to the query compilation step, and performs lightweight plaintext post-processing to emit outputs matching the corresponding \texttt{bedtools} command and flags (e.g., per-interval coverage summaries, window/intersect filtering, depth tables, or a single Jaccard summary).

\subsection{Server-side responsibilities}
\label{sec:system-overview-server}

\paragraph{Encrypted database hosting.}
The server stores the encrypted database as a collection of ciphertext chunks and provides indexed access to individual chunks for query processing.

\paragraph{Blind extraction kernel.}
For each extraction request, the server loads the referenced ciphertext chunk and executes the same fixed sequence of homomorphic operations: a slot rotation that moves the requested offset into a canonical slot, followed by multiplication with a plaintext mask that zeroes all other slots. The output is a ciphertext that contains exactly the requested scalar in a known slot, and this ciphertext is appended to the result stream.

\paragraph{Streaming execution and packaging.}
Because each extraction is independent, the server can process requests in a streaming manner (load chunk, rotate+mask, write output) and parallelize across CPU cores. The server returns the extracted ciphertexts together with minimal metadata needed for the client to decode and assemble them.

\subsection{End-to-End Query Workflow}
\label{sec:system-overview-workflow}

An end-to-end run consists of three stages.

\paragraph{Stage 1: One-time setup.}
The client builds the encrypted database from $B$, generates keys, and provisions the server with the encrypted database and evaluation keys. This stage is amortized across many queries and is repeated only when the database or cryptographic parameters change.

\paragraph{Stage 2: Per-query execution.}
Given $A$ and a selected command, the client emits a batched extraction request containing the physical addresses of all needed scalars. The server executes the rotate-and-mask kernel once per requested scalar and returns the extracted ciphertext bundle.

\paragraph{Stage 3: Client-side post-processing.}
The client decrypts the extracted ciphertexts, maps them back to the logical indices referenced by the query compilation step, and reconstructs the final \texttt{bedtools}-style output.

The subsequent sections describe the encrypted data layout and global-vector segmentation, the blind extraction protocol and its SEAL-level realization (rotations and masking), and the query compilation rules for each supported \texttt{bedtools} operation.

\section{Evaluation}
\label{sec:evaluation}

We evaluate \textsc{BEDCrypt} on a representative genomic interval analysis workload to measure end-to-end performance, ciphertext expansion overhead, and scalability characteristics. All experiments use the BFV homomorphic encryption scheme with polynomial modulus degree $N=8192$ and plaintext modulus of 30 bits, providing 128-bit security~\cite{seallib}. The client and server executables run on a macOS workstation with 12 CPU cores, and all measurements reflect wall-clock time, peak resident set size (RSS), and I/O bandwidth as captured by the benchmark instrumentation described in Section~\ref{sec:implementation}.

\subsection{Experimental Setup}
\label{sec:eval-setup-perf}

\paragraph{Dataset.}
We generated a synthetic BED workload consisting of a query file $A$ and a database file $B$ (10.86 MB total plaintext) on chr1 with coordinate space $L=100{,}000{,}000$ bases. This dataset is intended to stress-test the system with a realistic interval-analytics workload while remaining small enough for rapid iteration during development.

\paragraph{Measured operations.}
We benchmark five core \texttt{bedtools}~\cite{quinlan2010bedtools}-equivalent operations: \textbf{Coverage} (overlap counts and base-pair coverage per interval), \textbf{Depth} (per-base depth values within query intervals), \textbf{Window} (proximity-based overlap detection with configurable window size $w=5$, tested with \texttt{-c}, \texttt{-u}, and \texttt{-v} modes), \textbf{Intersect} (set-based intersection with \texttt{-u} and \texttt{-v} flags), and \textbf{Jaccard} (statistical similarity metric). Each operation exercises a different query compilation pattern and accesses distinct segments of the encrypted global database (Fig.~\ref{fig:system-design}B).

\paragraph{Metrics.}
For each operation, we report:
\begin{itemize}
    \item \textbf{Encryption phase:} Wall-clock time, peak RAM usage, and total I/O (ciphertext database size, representing client upload bandwidth in a network deployment).
    \item \textbf{Server compute phase:} Wall-clock time, peak RAM usage, and output I/O (size of extracted ciphertexts returned to client, representing client download bandwidth).
    \item \textbf{Decryption phase:} Wall-clock time and peak RAM usage. No additional I/O is reported, as bandwidth is accounted in the server phase.
\end{itemize}

\subsection{End-to-End Performance Results}
\label{sec:eval-perf-main}

Table~\ref{tab:bench-metrics} summarizes the measured time, peak RAM, and I/O per phase for each operation on the 10.86~MB dataset.

\paragraph{Noise management and feasibility.}
We use BFV parameters at polynomial modulus degree $N=8192$ with coefficient modulus close to 218 bits (SEAL \texttt{BFVDefault}), which is consistent with 128-bit security targets used in standard HE parameter selection guidance. The extraction circuit is intentionally shallow---one ciphertext rotation followed by one plaintext multiplication (masking)---so noise growth is modest and does not require relinearization (no ciphertext--ciphertext multiplication) or bootstrapping. For this evaluation, we additionally instrumented the server to report the noise budget along the extraction circuit. Empirically, a server-side noise-budget simulation reports an initial budget of 136 bits, decreasing to 133 bits after rotation and 98 bits after masking, confirming a substantial remaining margin for correctness under our extraction kernel.

\paragraph{End-to-end analysis.}
The workload exhibits a clear separation between a one-time provisioning cost and per-query costs. Coverage includes the database build/encryption step (24.7~s and 24,420~MB upload), which dominates its end-to-end time (71.8~s) and total I/O. Once provisioned, subsequent operations reuse the encrypted database and are driven primarily by server processing and decryption latency.
Across reused queries, total time ranges from 26.5~s (Depth) to 44.9~s (Jaccard). Server processing times (16.3--34.7~s) exceed client-side decryption (10.0--10.2~s), indicating that the dominant per-query cost lies in server-side homomorphic evaluation and response generation rather than local decryption. Server peak RAM remains constant at 14~MB across all operations, consistent with a streaming execution model. Client peak RAM is highest for Coverage due to the one-time encryption phase (7.8~GB), and for Depth due to output reconstruction over dense per-base results (1.9~GB); other operations remain below 120~MB during decryption.

\begin{table}[t]
\centering
\caption{\textbf{Benchmark results (10.86~MB dataset).} Measured wall-clock time (s), peak RAM (MB), and I/O (MB) per phase for each supported operation. ``Encryption'' is a one-time cost for provisioning the encrypted database and is reused across subsequent queries.}
\label{tab:bench-metrics}
\begin{tabular}{@{}llrrr@{}}
\toprule
\textbf{Operation} & \textbf{Phase} & \textbf{Time (s)} & \textbf{RAM (MB)} & \textbf{I/O (MB)} \\
\midrule
\multirow{3}{*}{Coverage}
  & Encryption & 24.7 & 7,846 & 24,420 (upload) \\
  & Server processing & 36.9 & 14 & 24,420 (download) \\
  & Decryption & 10.3 & 117 & 10 (results) \\
  \cmidrule(lr){2-5}
  & \textbf{Total} & \textbf{71.8} & \textbf{7,977} & \textbf{48,850} \\
\midrule
\multirow{3}{*}{Depth}
  & Encryption (reused) & --- & 6,488 & 336 \\
  & Server processing & 16.3 & 14 & - \\
  & Decryption & 10.2 & 1,871 & 336 (results) \\
  \cmidrule(lr){2-5}
  & \textbf{Total} & \textbf{26.5} & \textbf{8,373} & \textbf{672} \\
\midrule
\multirow{3}{*}{Window (\texttt{-c})}
  & Encryption (reused) & --- & 7,428 & - \\
  & Server processing & 19.4 & 14 & - \\
  & Decryption & 10.0 & 91 & 6 \\
  \cmidrule(lr){2-5}
  & \textbf{Total} & \textbf{29.5} & \textbf{7,533} & \textbf{6} \\
\midrule
\multirow{1}{*}{Window (\texttt{-u})} & Decryption (reused server) & 10.2 & 88 & 1 \\
\multirow{1}{*}{Window (\texttt{-v})} & Decryption (reused server) & 10.2 & 88 & 4 \\
\midrule
\multirow{3}{*}{Intersect (\texttt{-u})}
  & Encryption (reused) & --- & 6,637 & 1 \\
  & Server processing & 27.6 & 14 & - \\
  & Decryption & 10.1 & 90 & 1 \\
  \cmidrule(lr){2-5}
  & \textbf{Total} & \textbf{37.6} & \textbf{6,741} & \textbf{2} \\
\multirow{3}{*}{Intersect (\texttt{-v})}
  & Encryption (reused) & --- & 6,678 & --- \\
  & Server processing & 26.8 & 14 & --- \\
  & Decryption & 10.2 & 90 & 4 \\
  \cmidrule(lr){2-5}
  & \textbf{Total} & \textbf{37.0} & \textbf{6,782} & \textbf{4} \\
\midrule
\multirow{3}{*}{Jaccard}
  & Encryption (reused) & --- & 6,413 & - \\
  & Server processing & 34.7 & 14 & - \\
  & Decryption & 10.1 & 100 & - \\
  \cmidrule(lr){2-5}
  & \textbf{Total} & \textbf{44.9} & \textbf{6,527} & \textbf{-} \\
\bottomrule
\end{tabular}
\end{table}

\section{Discussion}
\label{sec:discussion}

In this work, we achieve privacy-preserving \texttt{BEDTools}~\cite{quinlan2010bedtools}-style genome interval analytics by letting an untrusted server operate only on an encrypted, reusable interval database and return encrypted scalars that the client reconstructs into standard outputs without disclosing plaintext loci.
Our key contribution is that core interval primitives (coverage summaries, intersections, window/proximity queries, and set-similarity via Jaccard) can be supported without genome-wide encrypted scans, which are typically the dominant barrier to practicality.
A central design decision is to answer queries via \emph{blind extraction} from permuted, SIMD-packed ciphertext chunks---so server work scales with the number of requested boundary values---rather than evaluating interval logic over an encrypted genome-wide representation or deploying heavier fully oblivious access mechanisms.
This ``extract-only-what-you-need'' execution model matches how interval tools are used in practice---many analyses request a relatively small set of boundary values per query interval---and therefore naturally targets settings where sensitive cohorts or proprietary tracks must be analyzed on shared compute or cloud infrastructure.

The discussion of privacy should be framed around what is and is not protected.
Under standard BFV IND-CPA security, the hosted database contents remain confidential, and the server cannot distinguish one depth/coverage value from another; additionally, the client-secret permutation prevents the server from directly mapping a requested chunk back to a genomic coordinate range.
At the same time, the system intentionally accepts access-pattern leakage at the level of which \emph{permuted} chunks and offsets are touched, and the total number of extractions per session; in a genomics setting, this motivates careful operational guidance (e.g., batching, query normalization, and avoiding highly distinctive repetitive queries) and provides a clear basis for future extensions that add stronger obliviousness when needed.

Finally, it is useful to consider how \textsc{bedCrypt} could fit into real analysis pipelines and what limitations should shape future work.
The current approach is most compelling for repeated querying against a stable reference track (or collection of tracks) where the one-time preprocessing/encryption cost can be amortized, and where returning compact summaries is sufficient for downstream steps.
Challenging directions include supporting richer multi-track operations, scaling across whole genomes and many samples without overwhelming key material or bandwidth, and hardening against active/malicious behavior (e.g., incorrect server responses) through lightweight verification.
Overall, \textsc{bedCrypt} bridges the gap between local-only analysis and fully oblivious protocols by providing strong confidentiality for interval data with a familiar, \texttt{bedtools}-style user experience and adequate performance for realistic workloads.

\section*{Code availability}
The code used in this study is available at \url{https://github.com/Georgakopoulos-Soares-lab/bedCrypt}.
``

\section*{Acknowledgements}
We thank Ioannis Mouratidis for helpful discussions and support.

\section*{Funding}
Research reported in this publication was supported by the National Institute of General Medical Sciences of the National Institutes of Health under award number R35GM155468.

\end{document}